1

# Quantitative Methods for Optimizing Patient Outcomes in Liver Transplantation

Raja Al-Bahou B.S.[1], Julia Bruner M.S.[1], Dr. Helen Moore Ph.D.[2], Dr. Ali Zarrinpar M.D, Ph.D.[1]*

*Affiliations:*

[1]Department of Surgery, University of Florida College of Medicine, Gainesville, Florida, USA

[2]Department of Medicine, University of Florida College of Medicine, Gainesville, Florida, USA

*Corresponding author, ali.zarrinpar@surgery.ufl.edu.**Visual Abstract:**

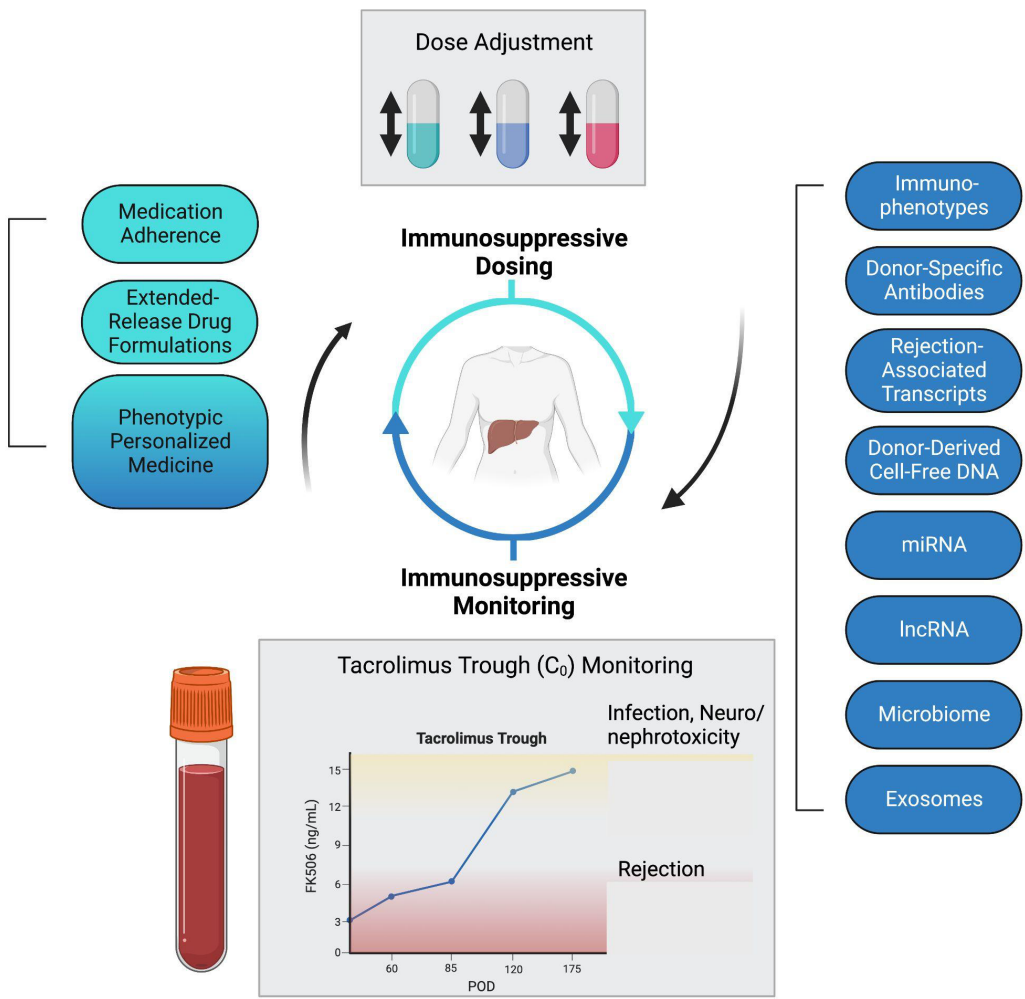

**Abstract:** Liver transplantation continues to be the gold standard for treating patients with end-stage liver diseases. However, despite the huge success of liver transplantation in improving patient outcomes, long term graft survival continues to be a major problem. The current clinical practice in the management of liver transplant patients is centered around immunosuppressive multidrug regimens. The goal of immunosuppression is reducing graft rejection. However, given the narrow therapeutic windows of these drugs, avoiding over-immunosuppression and under-immunosuppression continues to be a clinical challenge. Patients who are over-immunosuppressed are at higher risk for developing opportunistic infections; under-immunosuppressed patients may suffer graft rejection. Current dosing practices depend highly on physician experience and clinical judgment as well as dosing guidelines that are based on population averages describing the typical drug response. However, with a narrow therapeutic window, physicians tend to over or undershoot drug concentration, which increases the patient's time outside of the therapeutic range. Also, individual variability and pharmacogenomic polymorphism introduce additional levels of complexity to following dosing guidelines based on population averages. Thus, this has increased the need for a novel method to optimize dosing on an individualized level. Current research has been focusing on phenotypic personalized medicine as a novel approach in the optimization of immunosuppression, a regressional math modeling focusing on individual patient dose and response using specific markers like transaminases. While biomarkers like ALT and AST are typically used as a measurement of liver injury, some groups have looked into using cfDNA as a potential marker for measuring graft response to drug doses. Although our understanding of immunosuppression and liver transplantation is increasing rapidly, we still have miles to go. A prospective area of study includes the development of a mechanistic computational math modeling for optimizing immunosuppression to improve patient outcomes and increase long-term graft survival by exploring the intricate immune/drug interactions to help us further our understanding and management of medical problems like transplants, autoimmunity, and cancer therapy. Thus, by increasing long-term graft survival, the need for redo transplants will decrease, which will free up organs and potentially help with the organ shortage problem promoting equity and equal opportunity for transplants, as well as decreasing the medical costs associated with additional testing and hospital admissions. Although long-term graft survival remains challenging, computational and quantitative methods have led to significant improvements. In this article, we review recent advances and remaining opportunities. We focus on the following topics: donor organ availability and allocation with a focus on equity, monitoring of patient and graft health, and optimization of immunosuppression dosing.

**Keywords:** liver transplant, immunosuppression, tacrolimus,

**Introduction**

Advances in computational and quantitative methods and capabilities have led to increased understanding of complex biological and clinical problems. These advances are also beginning to have an effect on the delivery of care. Organ transplantation has greatly benefited from computational approaches and stands to benefit even more from future developments. This article will review recent advances in computation and quantitative methods applied to liver transplantation.

Liver transplantation is a life-saving procedure for patients with end-stage liver disease. Immunosuppression is a critical component of post-transplant management to prevent rejection and achieve long-term graft survival. The proper dosing of immunosuppressive medications is crucial, as underdosing can lead to graft rejection, while overdosing can increase the risk of infection. Because the level of immunosuppression is not yet a quantifiable measure and due to the large inter- and intra-individual variabilities in immunosuppression medication needs, individualized, dynamic dosing of immunosuppressive medications is critical to achieving optimal patient outcomes in both the short and long term. However, it is important to establish robust and reproducible methodologies to monitor patient immunosuppression levels and graft health prior to adjusting or optimizing care regimens.

The monitoring and optimization of immunosuppression are complex processes that require careful consideration of patient characteristics, medication interactions, and individual response to therapy. Various protocols and pharmacogenomics informed algorithms have been developed to improve immunosuppressive medication dosing for different patient populations, including a phenotypic personalized medicine (PPM) approach. Some have shown promise in improving individual dosing response and minimizing the risk of adverse effects. Combination therapy with immunosuppressive agents is often necessary to achieve optimal immunosuppression. However, the selection and dosing of combination therapy are challenging due to the potential for drug interactions and toxicity. Therefore, optimizing combination therapy requires careful monitoring and adjustment to achieve the best possible outcomes.

Finally, as organ shortage is a significant issue in all transplantation, including liver transplantation, there is a need to improve graft allocation and utilization while ensuring equity in access to organs. Advances in computation and quantitative methodology have provided valuable tools for addressing these complex issues. For example, machine learning algorithms have been developed to predict patient and graft survival and to optimize organ allocation to improve outcomes and equity in organ distribution.

In the review below, we will summarize these topics. The monitoring and optimization of immunosuppression are critical components of patient care in liver transplantation, and advances in quantitative methods have improved our understanding of these processes. Combining precision dosing and individualized combination therapy can lead to optimal outcomes, while machine learning algorithms can aid in improving graft allocation and utilization. By applying quantitative methods to the field of liver transplantation, we can continue to improve patient outcomes and address the challenges of organ shortage and equitable access to transplantation.

**Optimizing Liver Allocation and Utilization**

In recent years, expansion of research and knowledge in the field of transplantation has tremendously increased understanding of liver transplants, making them the standard treatment for patients with end-stage liver disease. This increased understanding of the complex biology of liver transplants, surgical techniques, pre-, peri-, and post-operative clinical management of allograft organs, immunology, and immunosuppressive therapy has expanded the indications for liver transplants to encompass more end-stage liver failure etiologies like acute and chronic liver failures, cirrhosis, several metabolic disorders, and select liver malignancies. However, this expanded usage contributed to a major problem facing the field of liver transplantation, namely, organ shortage. Organ shortage continues to be a major limitation for all solid organ transplants and a leading burden when it comes to liver transplant wait list mortality. Thus, improving upon current methods of liver allograft allocation and utilization is a necessity not only to expand the donor liver pools, but also to ensure equity across liver transplants opportunities and access to donor organs.

Over the years, many changes have been made to the liver transplant organ allocation and utilization policies with the sole goal of decreasing waitlist mortality by ensuring a fair and equitable distribution of the limited available liver organ pool. Traditionally, and in response to the increased demand of liver organs compared to the liver organ pools, the United Network for Organ Sharing (UNOS) followed a 'sickest first' allocation policy for liver organs in the United States, which based its allocation on medical urgency, where inpatients were given priority over outpatients with ICU admitted patients given the highest priority, as well as accumulated waitlist time[1]. However, as the number of patients requiring liver transplant increased, the UNOS status for organ allocation became less relevant compared to the accumulated waitlist time. Thus, hospitals started listing patients earlier to maximize their chances of receiving a donated organ. In response, UNOS adopted minimal listing criteria as well as an increased severity grading. The listing criteria were based on what was known as Child–Turcotte–Pugh score, which was composed of five components including bilirubin, international normalized ratio (INR), albumin level, encephalopathy, and ascites. However, considering the amount of ascites and the level of encephalopathy are two subjective parameters, a more objective and quantifiable allocation system was needed to ensure fair and equitable distribution of the available organ pool.

Thus, UNOS supported the replacement of the Child–Turcotte–Pugh classification with the Mayo Clinic developed Model for End-Stage Liver Disease (MELD) scoring system as a basis for liver organ allocation, which was adopted in the United States in 2002. The MELD score is based on three objective clinical parameters in its calculation, which include serum bilirubin, serum creatinine, and INR for prothrombin time, as a predictor for 90-day mortality in patients with cirrhosis, and is widely used now to stratify patients based on their need for a transplant. In 2016, the MELD score was adjusted to include serum sodium in its calculation as a better predictor of 90-day mortality[1]. More recently, MELD 3.0, an optimized version of MELDNa, was shown to be a better short term mortality predictor compared to MELDNa while also addressing determinants of wait list outcomes like sex disparity[2]. The creation and adoption of the MELD scoring system as the basis for liver organ allocation and utilization reduced the transplant waitlist mortality by prioritizing sicker patients without impacting their post transplant outcomes[1].

On the other hand, to ensure a fair and equitable distribution of resources, the Organ Procurement Organizations (OPOs), non-profit organizations that are responsible for recovering the donated organ from the donors, and are especially trained to deal with the donor families, match the donor organs to the waitlist recipients through the Organ Procurement and Transplantation Network (OPTN), which is directly managed by UNOS. Currently, there are 56 OPOs in the United states responsible for carrying out this task[3]. Previously, every OPO was tasked with serving their specific Donor Service Area (DSA), which combined to make eleven UNOS regions[1]. Prior to the implementation of acuity circles distribution policy, the distribution of donor livers within a region was limited by their donation service area (DSA). For instance, when a donor liver became available, it was first offered to its local OPO within a specific DSA and matched to the patient with the highest MELD score. Only when a liver was not accepted by the local OPOs following severity level, was it offered to the regional

OPOs and then nationally. This geographical limitation gave rise to a significant concern regarding fairness in the allocation process of liver organs, which accounts to differences in population size and demographics within a particular region leading to geographical disparities and by extension inequity in the access to liver transplantation.

Recently, UNOS adopted different criteria in response to persistent significant variability in organ allocation and distribution despite the adoption of different policies like Regional Share for Status 1, Regional Share 35, and National Share 15, in an effort to decrease waitlist mortality and geographical variability within the previous local-regional-national distribution system. Such efforts lead to the elimination of the DSAs and UNOS regions in liver organ distribution, and the implementation of the new Acuity Circles distribution system, which operates on the basis of concentric geographic circles around the donor site hospital. For instance, when an organ becomes available, it will be first offered to Status 1 patients within 500 nautical miles (NM) of the donor hospital site. Then, it will be offered to higher severity patients with a minimum MELD of 37 within 150 NM, then 250 NM, and finally 500 NM. Only when an organ was not accepted for any of these patients, will it be offered to patients with decreasing MELD scores of 33, 29, and 15 in expanding concentric circles at each of these MELD scores until being allocated nationally and finally with less severe patients with MELD score lower than 15[1]. The acuity circles distribution system was accepted in 2018 and implemented in 2020. While this new system helps minimize geographical disparities in liver organ allocation by prioritizing higher MELD patients with increasing radius, concerns about distance traveled and organ viability is still in question. An OPTN report of two year monitoring of liver and intestine acuity circle allocation stated a significant transplant rate increase for liver-alone candidates with MELD or PELD scores of 15 and lower and 29 and higher as well as for Status 1A/1B candidates. Also, it showed an increase in distances between donor hospital and transplant program for deceased donor liver-alone recipients, as well as a slightly increased discard rate and a decreased liver utilization rate from before policy implementation compared to post-policy implementation[4]. In addition, the newly implemented acuity circles policy raised concerns of fairness of organ redistribution. It was argued that states with longer waitlists stand to benefit the most from the recent change even when they performed the worst in organ procurement, which brings the issue of resource redistribution from states with better organ procurement rates and higher disadvantaged population in terms of access to care and insurance. Thus, taking resources away from disadvantaged states and increasing the inequity in organ allocation for low socioeconomic patients or patients without access to proper healthcare[5].

As organ shortage continues to be a significant limitation in all solid organ transplants, including liver, there is a need to improve graft allocation and utilization while ensuring equity in access to organs. Advances in computation and quantitative methods have provided valuable tools for addressing these complex issues. For example, machine learning algorithms have been proposed and developed to predict patient and graft survival and to optimize organ allocation to improve outcomes and equity in organ distribution[6,7].

**Immunosuppression Monitoring and Diagnostics**

Given the narrow therapeutic index of immunosuppressants such as tacrolimus, consistent monitoring is necessary to maximize efficacy and minimize toxicity. The standard of care for immunosuppressant drug level monitoring after liver transplantation is whole blood measurement of tacrolimus pre-dose or trough ($C_0$) concentration. Equivalent trough measurements are available for cyclosporine but less commonly utilized, matching clinical preference for tacrolimus[8].

Clinicians use blood drug concentration labs to inform immunosuppression management decisions, which aim to maintain immunosuppressants within target therapeutic concentration windows. However, high inter- and intra-patient variability in response to immunosuppressants, particularly CNIs, complicate modulation. Due to such variability in response, immunosuppressive needs indicated by drug level monitoring may not accurately reflect the dose requirements of the patient. As a result, patient blood drug concentrations frequently deviate from the recommended therapeutic concentration ranges targeted by managing clinicians. Both intra-patient variability and deviation from targeted windows has been associated with increased risk of nephrotoxicity, infection, and rejection as well as poor long-term outcomes after liver transplantation[9]. Hence, existing immunosuppressant monitoring practices lack capacity to account for variability. Supplementary or novel monitoring techniques are necessary to reduce clinician prescriptive guesswork and improve long-term patient outcomes.

In the long-term setting, a preliminary prospective study by Leino et al. indicates that adherence may be a significant factor contributing to intrapatient variability in response to immunosuppression[10]. Quantitative analysis of daily tacrolimus trough levels in a small cohort of adherent renal and liver transplant recipients was used to calculate weekly median coefficient of variation (CV). Median CV for liver transplant recipients was 15.2%, falling significantly below 30% within subject CV typically exhibited by highly variable drugs, though this refers to AUC or $C_{max}$ rather than $C_0$ pharmacokinetics. Given that tacrolimus $C_0$ is the standard of care and well correlated to AUC, this result is significant. Despite limitations in sample size, the prospective nature and frequency of collections indicate promise for further investigation. Similarly, a recent multicenter trial by Melilli et al. evaluating use of a medication adherence smart-phone application found that patients who took more than 20% of their immunosuppressant doses out of window (>2 hours from scheduled dose) in the 6 days prior to blood trough level assessments had significantly higher intra-patient variability (17% vs. 29%) and a significantly greater number of dose adjustments[11]. Interventions increasing strict adherence to long-term immunosuppression protocols may help improve intra-patient variability to complement existing drug monitoring techniques.

Several long-term studies indicate switching patients from twice-daily to a once-daily extended-release form of tacrolimus reduced patient burden and increased adherence[12–16]. In a decade-long follow up study by Toti et al. switching liver transplant recipients to an extended-release tacrolimus not only reduced nonadherence by 53.3% but increased renal function and self-reported patient quality of life[14]. Increased renal function (eGFR) is not commonly reported by studies evaluating the switch to once-daily tacrolimus formulations, however duration of follow up may be implicated. In several studies, prolonged-release tacrolimus formulations were associated with decreases in interpatient variability, and were also related to a reduced need for dose adjustments to maintain a therapeutic tacrolimus $C_0$.[13,16]. Implementation of strategies to support better medication adherence may support existing quantitative drug level monitoring methods by decreasing intra-patient variability to promote predictability of tacrolimus troughs.

Aside from direct blood drug concentration measurements, emerging quantitative methods aim to monitor immunosuppression indirectly through its effect on the immune system. ImmuKnow Immune Cell Function Assay by Viracor (previously Cylex) measures adenosine triphosphate (ATP) production following phytohemagglutinin (PHA) mitogen induced CD4+ T-cell stimulation using whole blood. Increased ImmuKnow scores (ATP concentrations) imply T-cell activation and may indicate immunosuppressive need or impending rejection events. However, clinical validation studies describe significant variation in results. A meta-analysis of 6 total studies using ImmuKnow to assess rejection risk (n=1), infection risk (n=1), or both infection and infection risk (n=4) found that for prediction of rejection, ImmuKnow had pooled diagnostic odds ratio (DOR) of 8.8, sensitivity of 65.6%, and specificity of 80.4%. However, DOR ranged from 1.6-24.5, sensitivity from 9.1%-85.7%, and specificity from 76.4%-98.7%, demonstrating substantial variation. For diagnosis of infection ImmuKnow demonstrated pooled DOR of 14.7, sensitivity of 83.8% and a specificity of 75.3% Variations were lower for infection studies with DOR ranging from 11.4 to 85.7, sensitivity from 79.2% to 96.8% and specificity from 69.6% to 79.4%, respectively[17]. Challenges with reliability of the ImmuKnow assay results may stem from T-cell isolation or stimulation steps. T-cell isolation requires mechanical or chemical separation techniques which often damage the sample population. The stimulation step uses PHA mitogen as a non-specific antigen to stimulate adaptive, but not innate immunity which may also contribute to infection and rejection processes.

Despite showing little association with rejection or long-term outcome in initial studies, more recent literature demonstrates that de novo donor-specific antibody (dn-DSA) formation is associated with reduced graft and patient survival [12]. A systematic meta-analysis of 15 studies studying the relationship between dn-DSA development and rejection (n=10) or allograft loss (n=5) revealed that in the diagnosis of rejection, dn-DSA formation DOR was 6.43 (CI 3.17–13.04) for DSA-positive patients relative to the DSA-negative patients. The meta-analysis by Bayzaei et al. also found increased sensitivity to predict rejection for living donor liver transplantation recipients (DOR = 15.44 with 95% CI 6.32–37.74) compared to deceased donor recipients (DOR = 1.75; 95% CI 0.76–4.02). Rate of allograft loss was not statistically different between DSA-positive patients and DSA-negative patients. Quantitation below mean fluorescence intensity MFI of 3000 displayed consistent results but above 3000 MFI results were insignificant[12]. This threshold has been further implicated in studies examining use of dn-DSA for risk detection in other solid organ transplantation fields [13, 14].

Quantification of donor-derived cell-free DNA (dd-cfDNA) has also been used to predict rejection in liver transplantation. Utility of dd-cfDNA as a biomarker of rejection was previously discovered and applied in renal transplantation but recent studies examine its translation to the liver transplant setting. In the 2 initial investigations quantitatively assessing diagnostic capacity of dd-cdDNA in liver transplantation, Schütz et al. found an AUC of 0.97 for biopsy proven acute rejection and Goh et al. found an AUC of 0.97 for treated biopsy proven acute rejection with rejection activity index greater than 3. Despite matching AUC's, Schütz et al. used a threshold of 10% dd-cfDNA fraction to indicate rejection while Goh et al used absolute quantification with a threshold of 898 copies/mL to indicate rejection. Both studies employed digital droplet PCR to complete quantification[18,19]. Levitsky et al. confirmed that elevated dd-cfDNA fraction is associated with rejection and compared its diagnostic capacity to standard of care liver function tests (LFTs)[20]. Using a threshold of 5.3%, the AUC of dd-cfDNA to distinguish rejection from healthy post-transplant status was 0.95. However, the AUC for standard of care alanine transaminase (ALT) was 0.99 and combined with dd-cfDNA, 1.00. Distinguishing acute rejection (dd-cfDNA>20.4% ) from acute dysfunction without rejection, the AUC for dd-cfDNA at threshold 20.4% was 0.71, slightly greater than ALT alone, which demonstrated AUC of 0.69. Again, combination of dd-cfDNA and ALT provided only marginal improvement, with AUC of 0.72. When both non-rejection patient sets, acute dysfunction without rejection and stable post-transplant status, were grouped, the AUC of dd-cfDNA to distinguish acute rejection from no rejection was 0.85 at threshold 15%. Both alone and combined with dd-cfDNA, the AUC for ALT was 0.86. In each case, dd-cfDNA did not show significant improvement over standard of care LFTs for diagnosis of rejection. Given dd-cfDNA quantification is considerably more costly and time consuming, clinical utility remains limited.

Another approach to developing molecular diagnostic tools by Madill-Thomsen et al. used RNA microarray of liver biopsies to monitor expression of 417 rejection-associated transcripts derived and annotated in renal transplantation. Machine learning including unsupervised archetypal analysis (AA) and principal component analysis (PCA) were used to stratify outcomes based on expression profiles. Similar methods were used previously by this group to develop the Molecular Microscope® Diagnostic System (MMDx) for kidney, heart, and lung transplantation. Using 253 liver biopsy samples from 10 different centers, unsupervised archetypal analysis identified 4 groups: healthy patients ($R_{normal}$, N = 129), T cell–mediated rejection ($R2_{TCMR}$, N = 37), early injury ($R3_{injury}$, N = 61), and fibrosis ($R4_{late}$, N = 8). While an AMBR signature was identified in previous analysis, no such phenotype was revealed in liver transplantation[21]. In the future, molecular diagnosis strategies may address limitations of biopsy including sampling error and lack of reproducibility. However, because biopsy is both invasive and costly most molecular diagnosis investigations aim to create diagnostic tools reliant on biomarkers observable in blood rather than directly from tissue samples.

Development and optimization of novel biomarkers of rejection may help enhance existing therapeutic drug monitoring practices. Greater capacity to predictively diagnose adverse outcomes such as rejection may aid clinicians to anticipate changes in immunosuppressive needs and adjust medication prophylactically. Modulating immunosuppression before rejection or other phenotypes clinically manifest may better protect both long and short-term patient outcomes after liver transplantation. However, prospective drug modulation interventions based on these potential biomarkers have not yet been completed. In the future, emerging targets for monitoring will need to be evaluated for their capacity to guide clinician immunosuppression management in addition to their diagnostic capacity. Other likely directions of future exploration include novel monitoring targets including non-coding RNAs such as microRNAs (miRNA) and long non-coding RNAs (lncRNA), microbiome signatures, and the role of extracellular vesicles including exosomes in communication between the innate and adaptive immune system. Initial investigations in these areas have identified specific targets which differ significantly between patients in acute rejection and patients with stable status, however, the diagnostic capacity of such targets has not yet been quantitatively evaluated, validated, or compared to existing markers.

**Immunosuppression Dosing**

Drug regimens can make a tremendous difference in patient outcomes. One historical example is the improved survival rates in childhood acute lymphoblastic leukemia (ALL) achieved over decades of experimentation with chemotherapy drug regimens. Inaba and Pui (2021) compared survival rates of children with ALL over the years at a single treatment center (St. Jude Children's Research Hospital)[22]. In 1962, ALL 10-year survival rates at St. Jude were 11.1% (+/-3.2%). In 2006, ALL 10-year survival rates at St. Jude were over 91%. Much of the improvement over those decades was due to improved regimens, as most conventional chemotherapies were approved in this setting before 1980, and no new therapies were approved between 1979 and 1994. The large improvements in 10-year survival can be seen in the histogram in Fig. 1. These improvements obtained through experimentation required hundreds of patients and many decades.

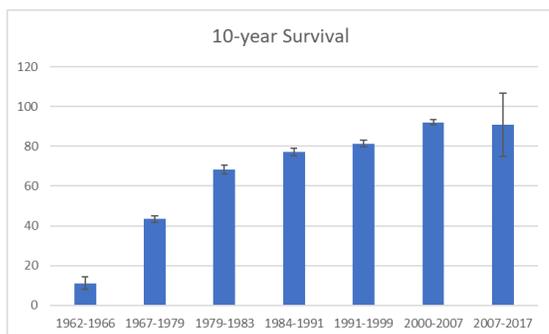

**Fig. 1** Ten-year overall survival in pediatric patients with acute lymphoblastic leukemia (ALL) treated in studies at St. Jude Children's Research Hospital. Height of blue bars shows the percentage of patients surviving ten years. Standard deviations for each cohort are shown in black bars. Data from Inaba and Pui[22].

Another large clinical difference between regimens was found in the setting of breast cancer. Alternating schedules of multiple therapies were hypothesized to be more effective and cause less drug resistance than a sequential schedule. This hypothesis of Goldie and Coldman[23] was refuted in a study by Bonadonna et al. of hundreds of patients[24]. The Norton and Simon hypothesis[25] proposed that dose-dense treatment schedules would be more effective than standard schedules. This hypothesis was proven correct, in a study that required thousands of patients over years[26]. Although clinical experimentation can resolve questions about dosing regimen efficacy, it can take long periods of time and large numbers of patients.

In the remainder of this section, we will examine other methods for optimizing drug regimens to improve patient outcomes in the setting of liver transplant. We will focus on the use of modeling, both empirical and mechanistic, to speed this process of improving drug regimens and patient outcomes.

Current practice for post-transplant immunosuppression is specific to the organ transplanted, as well as the center where the transplant was performed. Dose-adjustment protocols consider concentrations of immune-suppressing drugs and markers of transplant damage. After liver transplantation, a mainstay of the immunosuppression combination regimen is the macrolide compound tacrolimus[27]. Other drugs usually used in combination with tacrolimus include one of four steroid formulations and a drug such as mycophenolate mofetil. Tacrolimus has a markedly small therapeutic index: concentrations that are too low can trigger graft rejection, concentrations that are too high can allow opportunistic infections. This is a challenge, because both over- and under-suppression of immune function appear to cause large fractions of transplant patient deaths[28]. The concentrations of tacrolimus are also highly variable, both between individuals, and within individuals over time. For these reasons, regimen optimization in liver transplant patients has necessarily focused on tacrolimus, and has included therapeutic drug monitoring (TDM)[29]. TDM allows for dosing adjustments to be made for each patient individually, in real time.

Tacrolimus dosing generally needs to take into account various use settings. For example, patients transplanted because of autoimmune liver disease require higher levels of tacrolimus. Patients transplanted for alcohol-associated liver disease or hemochromatosis can be treated with lower levels[27]. Adjustment of tacrolimus doses based on genotype is another important consideration, and an area of active exploration. CYP3A5 genotype has been found to affect the drug concentrations of tacrolimus in liver transplant patients, though CYP3A4 and ABCB1 do not seem to have such an effect[30]. CYP3A5 can be expressed in both the liver and the organ, so both the donor and the recipient CYP3A5 expression can play a role. The benefit of such genotyping is to better understand the expected pharmacokinetics of tacrolimus before dosing, and to adjust the doses accordingly in advance, in a precision medicine approach. Patients with CYP3A5 genotype require approximately 50% higher tacrolimus dose to achieve target therapeutic ranges of tacrolimus, compared to patients without this genotype[31].

Mathematical and statistical modeling can help improve patient outcomes in settings like liver transplant, while reducing the time and number of patients that would be required to find better regimens experimentally. One traditional type of statistical/empirical modeling, population pharmacokinetics (PopPK) modeling, has been used extensively to model tacrolimus concentrations in liver transplant, and has been refined over the years. A review of sixteen of these PopPK models was published in 2020 by Cai et al[32]. Some of the most common significant covariates included in these models are post-operative days, hematocrit, and total bilirubin. Physiologically-based pharmacokinetic (PBPK) models have been used to obtain more explanation of the variability in tacrolimus concentrations in liver patients. For example, Gérard et al. found that the most influential covariates on tacrolimus trough concentrations were unbound drug fraction, intrinsic clearance, CYP3A5 genotype of the liver donor, and recipient body weight and hematocrit[33].

Shi et al. developed an empirical statistical model for post-liver transplant tacrolimus concentrations[34]. The model included covariates for donor and recipient genotype, and recipient weight and total bilirubin, and was validated clinically. Compared to physician-guided dose adjustments, the statistical model-guided dosing was more likely to achieve the target concentration range of 4–10 ng/mL of tacrolimus with more-individualized dosing (0.023-0.096 mg/kg/day for the model-guided dosing vs. 0.045-0.057 mg/kg/day for the physician-guided dosing). Additionally, significantly fewer later dose adjustments were needed for the model-guided group vs. the physician-guided group (2.75 +/- 2.01 vs. 6.05 +/- 3.35, p = 0.001).

In the early postoperative setting, intrapatient variability is hypothesized to depend on a variety of factors including MELD score and Child-Pugh grade, CYP3A5 and other drug metabolism or transport related genotypes, site specific practices, and environmental factors[9]. To account for these factors, we previously developed an empirical, mechanism-free approach that implicitly addresses the range of such variables to supplement existing immunosuppression monitoring and modulation practices. Such approaches have been proposed previously[6]. Our team developed a phenotypic personalized medicine (PPM) platform that uses each patient's clinical data to create a model of their dose-response relationship using quadratic or linear regression. The dose-response model is then used to calculate each patient's unique optimal dose. We demonstrated survival benefits for liver transplant patients using a quadratic regression PPM to guide immune-suppressing dosing in a pilot study[35]. In a phase 2 randomized clinical trial of 62 patients, dosing based on a PPM linear regression kept patients in range of a specified trough concentration for a higher percentage of time than the standard of care dosing did; it also decreased the length of hospital stays by 33%[36].

Beyond empirical and statistical modeling methods for determining dose adjustments, mechanistic modeling can provide dosing adjustments that further improve patient outcomes. One highly successful example is embodied in the artificial pancreas, which optimizes insulin dosing in Type I diabetics[37]. The use of a mechanistic mathematical model was critical to its success. Models that were not mechanistic and simply relied on current levels of glucose to make dose adjustments to insulin were not as successful, primarily due to the lag time in the glucose-insulin dynamics. The mechanistic model allowed for anticipation of the future dynamics in adjusting the current insulin dose. This approach is now being used more widely in a variety of applications, and could be used for the dosing of tacrolimus for liver transplant patients, as well. A mechanistic approach would allow more accurate individualization of tacrolimus dosing, resulting in longer time periods within specified concentration ranges.

**Discussion**

Although the current advances in the field of liver transplantation have revolutionized the clinical management and care for patients with end stage liver disease over the past decades, the field continues to face many challenges that limit its progress.

For instance, the increased demand of donated liver organ compared to the limited organ pool continues to be a major limitation in liver transplantation. As a result, the numbers of waitlisted patients continues to rise necessitating the need for a computational and quantifiable approach in ensuring equity in organ allocation and utilization. Also, given the scarcity of organ pools, there is a pressing motivation to address disparities based on race, socioeconomic status, and geography to ensure equitable access to organ transplantation. Over the years, many efforts have been made to try and address this problem moving away from subjective assessments like medical urgency and towards a more quantifiable approach like MELD scoring system. On the other hand, while most efforts have mainly looked at improving allocation systems, looking for ways to increase the organ pool is another side that must be considered in addressing this problem. For instance, requiring efforts to enhance deceased organ donation rates, promoting living donation programs, as well as exploring alternative methods. Besides, inclusion criteria for transplantation must consider factors beyond medical urgency, such as post-transplant success and quality of life. In addition, advancements in organ preservation technologies are necessary to optimize organ viability during transportation and improve utilization rates. Lastly, achieving more equitable distribution entails addressing geographic disparities and implementing objective allocation policies. Thus, by addressing these challenges, patient outcomes can be enhanced, and access to life-saving transplantation can be improved for all individuals in need.

Another issue facing the field of transplantation is the long term maintenance of patients on immunosuppression. Immunosuppression markers are indicators used to assess the level of immunosuppression in transplant recipients throughout their treatment. These markers help monitor the effectiveness of immunosuppressive therapies and aid in adjusting medication dosages. Common immunosuppression markers that are currently used clinically include transaminases, which indicate liver injury. Monitoring immunosuppression markers plays a crucial role in maintaining the delicate balance between preventing organ rejection and minimizing the risk of complications related to excessive immunosuppression. Better markers to detect impending risk for graft rejection or infection would help.

Additionally, from prior examples for other diseases, we know that improving dose regimens can make a great difference. In the liver transplant setting, much of the focus is on appropriately dosing the commonly-used immunosuppressant tacrolimus. Tacrolimus has a narrow therapeutic window, but also has large inter- and intra-patient variability. Dose recommendations have used covariates such as genotype, transplant reason, and other health status. Beyond this, many dose adjustments are physician-guided based on medical experience and population averages studies. There remains a need to establish evidence-based quantitative methods for tacrolimus dose adjustments, as increasing the time within target concentration ranges leads to better patient outcomes. Additional informative biomarkers could support TDM and improve quantitative methods.